\documentclass[fleqn,10pt]{wlscirep}
\usepackage[utf8]{inputenc}
\usepackage[T1]{fontenc}
\usepackage{lineno}
\usepackage[most]{tcolorbox}
\usepackage{capt-of}
\nolinenumbers

\usepackage[utf8]{inputenc} 
\usepackage[T1]{fontenc}

\UseRawInputEncoding

\usepackage{amsmath,amsfonts,bm}

\def\eqref#1{equation~\ref{#1}}

\def\1{\bm{1}}

\def\va{{\bm{a}}}

\def\vg{{\bm{g}}}

\def\vr{{\bm{r}}}
\def\vs{{\bm{s}}}

\def\vv{{\bm{v}}}

\def\vx{{\bm{x}}}

\def\vz{{\bm{z}}}

\def\evz{{z}}

\def\mB{{\bm{B}}}

\def\mI{{\bm{I}}}

\def\mM{{\bm{M}}}

\def\mS{{\bm{S}}}
\def\mT{{\bm{T}}}

\def\mV{{\bm{V}}}
\def\mW{{\bm{W}}}

\DeclareMathAlphabet{\mathsfit}{\encodingdefault}{\sfdefault}{m}{sl}
\SetMathAlphabet{\mathsfit}{bold}{\encodingdefault}{\sfdefault}{bx}{n}

\def\sA{{\mathbb{A}}}

\def\sX{{\mathbb{X}}}

\def\sZ{{\mathbb{Z}}}

\def\emS{{S}}
\def\emT{{T}}

\newcommand{\myequation}{\begin{equation}}
\newcommand{\myendequation}{\end{equation}}
\let\[\myequation
\let\]\myendequation

\newtcolorbox[auto counter]{boxy}[1][]{
  enhanced,
  breakable,
  fonttitle=\scshape,
  #1,
  before upper={\parindent15pt}
}

\title{How to build a cognitive map: insights from models of the hippocampal formation}

\author[1,2,$\dag$,*]{James C.R. Whittington}
\author[2,$\dag$]{David McCaffary}
\author[2]{Jacob J.W. Bakermans}
\author[2,3,4]{Timothy E.J. Behrens}
\affil[1]{Department of Applied Physics, Stanford University, Stanford, CA, USA}
\affil[2]{Wellcome Centre for Integrative Neuroimaging, University of Oxford, Oxford, OX3 9DU, UK}
\affil[3]{Wellcome Centre for Human Neuroimaging, University College London, London, WC1N 3AR, UK}
\affil[4]{Sainsbury Wellcome Centre for Neural Circuits and Behaviour, University College London, London W1T 4JG, UK}

\affil[*]{corresponding author(s): \texttt{jcrwhittington@gmail.com}}
\affil[$\dag$]{these authors contributed equally to this work}

\begin{abstract}

Learning and interpreting the structure of the environment is an innate feature of biological systems, and is integral to guiding flexible behaviours for evolutionary viability. The concept of a \textit{cognitive map} has emerged as one of the leading metaphors for these capacities, and unravelling the learning and neural representation of such a map has become a central focus of neuroscience. While experimentalists are providing a detailed picture of the neural substrate of cognitive maps in hippocampus and beyond, theorists have been busy building models to bridge the divide between neurons, computation, and behaviour. These models can account for a variety of known representations and neural phenomena, but often provide a differing understanding of not only the underlying principles of cognitive maps, but also the respective roles of hippocampus and cortex. In this Perspective, we bring many of these models into a common language, distil their underlying principles of constructing cognitive maps, provide novel (re)interpretations for neural phenomena, suggest how the principles can be extended to account for prefrontal cortex representations and, finally, speculate on the role of cognitive maps in higher cognitive capacities.

\end{abstract}
\begin{document}

\flushbottom
\maketitle

\thispagestyle{empty}

\section*{Introduction}

\begin{figure}[!b]
\centering
\includegraphics[width=0.96\linewidth]{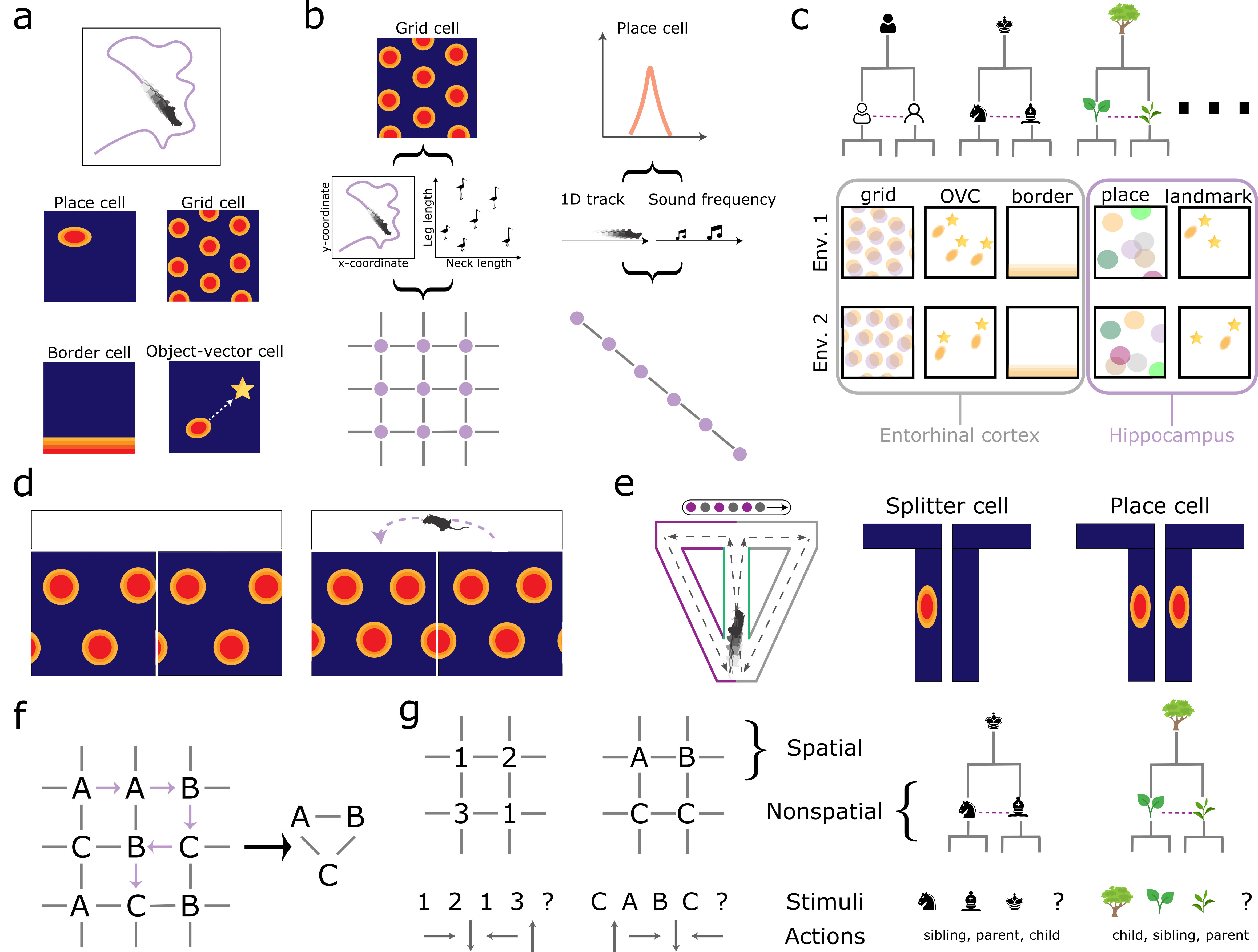}
\caption{\textbf{The cognitive mapping problem: generalisation and latent states.} \textbf{(a)} When navigating naturalistic environments, a range of cell representations are found in the cognitive map of the hippocampal-entorhinal system. Many of these representations were discovered in the context of the spatial cognitive map in rodents \cite{Hafting2005, OKeefe1971}. \textbf{(b)} Recent evidence, however, has implicated these same representations in the coding of abstract or conceptual spaces (e.g. `bird space' \cite{Constantinescu2016} or sound frequencies \cite{Aronov2017}), with subsequent theoretical accounts suggesting a single coding mechanism underlies both physical and conceptual spaces \cite{Behrens2018, Bellmund2018, Whittington2020}: understanding the relational structure (e.g. via \textit{graphs}) of the world. \textbf{(c)} Understanding relational structure abstractly affords generalisation, since sensory particularities can differ across instances of the same underlying structure \cite{Behrens2018}. Cell representations of the entorhinal cortex map relational spaces and \textit{generalise} across environments more so than hippocampal cells. \textbf{(d)} Since the sensory world is aliased, representations must be \textit{latent} (not a simple function of the current observation). Rodents exhibit latent state representations when they traverse two sensorially identical rooms \cite{Carpenter2015}. Initially, an identical grid cell code represents both rooms, though when the animal realises these two rooms are connected by a corridor, a \textit{global} grid cell code predominates; the latent representations separate out states with differing sensory futures. \textbf{(e)} Latent states are not just in space. In a T-maze alternation task \cite{Wood2000}, where rodents take alternating left and right turns (\( left \to right \to left \to right \cdots \)), in addition to spatial place cells, `splitter cell' representations form, which fire preferentially on left/right trials. These are non-spatial latent state representations, since they disambiguate the same spatial location (central trunk) depending on whether it is predicting `go left' or `go right'. \textbf{(f)} The aliasing problem in graphs: by representing state via an observation, these two graphs would appear the same. \textbf{(g)} Sequence prediction tasks are sufficient to learn latent state representations, since identical observations can have different neighbours. Sensory sequences, and the associated actions, can come from both space and non-space (e.g. families). Some sensory predictions can only be done by knowing (generalising) certain rules e.g. \texttt{North} + \texttt{East} + \texttt{South} + \texttt{West} = 0 or \texttt{Parent} + \texttt{Sibling} + \texttt{Niece} = 0.}
\label{fig:intro}
\end{figure}

Since the 1950s, the hippocampal formation has been implicated in numerous different functions, ranging from episodic memory to spatial and abstract cognition \cite{Scoville1957, Cohen1980, OKeefe1978, Hafting2005, Hassabis2007}. In this time, neuroscientists have attempted to characterise, and provide normative explanations for, the neural representations supporting such functions. Nowhere has this approach proved more fruitful than in the spatial domain, where a variety of cell types, including hippocampal place cells and entorhinal grid cells, provide an appealing neural instantiation of Tolman's (and Turner's) cognitive map \cite{Tolman1948, Turner1907, OKeefe1978, Hafting2005} (Figure \ref{fig:intro}a).

Cognitive maps were originally proposed as internal neural representations affording flexible behaviour, such as planning routes or taking never-before-seen shortcuts \cite{Tolman1948, Turner1907, Zanforlin1970}. More recent descriptions formalised this view with the key concept of \textbf{generalisation} \cite{OKeefe1978, Cohen1980, Behrens2018}. Here, the fundamental role of cognitive maps is to organise knowledge, facilitating generalisation of this knowledge to novel experiences, and thus enabling the rapid inference from sparse observations which characterises biological intelligence \cite{Humboldt1836, Tenenbaum2011a}. Psychologists have thought similarly, both with schemas \cite{Bartlett1932} (a mental framework for understanding new information), and with `learning to learn' \cite{Harlow1949a} (learning underlying rules of tasks that permit more efficient learning for each new task instantiation). While all these concepts are broad, encompassing domains from social to logical cognition \cite{Tolman1948}, most neural evidence for a cognitive map is grounded in studies of space \cite{OKeefe1978, Moser2017}.

Recent experimental results, however, increasingly suggest deep parallels between spatial cognition and abstract, non-spatial reasoning \cite{Behrens2018} (Figure \ref{fig:intro}b). For instance, hippocampal place cells, which fire with remarkable precision when the animal is in one location in space, also code for one `place' in sound frequencies \cite{Aronov2017} when sound frequency is an important component of the given task, or one `place' in abstract spaces mapped via value \cite{Knudsen2020} or integrated sensory evidence \cite{Nieh2021}. Similarly, the characteristic hexagonal firing pattern of entorhinal grid cells, discovered in the context of physical space \cite{Hafting2005}, is also found when animals navigate abstract spaces \cite{Constantinescu2016, Bongioanni2021, Bao2019, Park2020}. For example, in fMRI, human entorhinal cortex (and medial prefrontal cortex; mPFC) \cite{Constantinescu2016, Bao2019, Park2020} and monkey mPFC \cite{Bongioanni2021} display a hexagonally symmetric pattern when stimuli varying along two abstract dimensions (such as neck and leg length of birds \cite{Constantinescu2016}, odours \cite{Bao2019}, social hierarchies \cite{Park2020}, or reward probability and value \cite{Bongioanni2021}) are presented. These parallels in representation suggest the mechanism for constructing the spatial cognitive map might, in fact, be an instance of a more general coding mechanism capable of building abstract cognitive maps covering any domain.

This presents the exciting and novel opportunity to understand how the brain represents these apparently divergent domains of cognition in the same way. Developing such an understanding, however, requires a formalism connecting physical and abstract space \cite{Behrens2018}. In recent years, many models of the hippocampal formation have attempted to do this, providing explanations of neural data and offering falsifiable predictions. While greatly informative, these models differ in their focus and the language of their formalism, obscuring the overall direction and vast potential of this work. The aim of this Perspective is to clarify the common theory underlying these models \footnotemark, while providing novel results offering normative explanations for a range of old and new neural phenomena. We conclude with a prospective account, speculating how far these models might take us into understanding the neural representations of higher-order cognitive domains, such as language, logical operators, and mathematics, thereby providing a pathway towards cognitive maps as Tolman envisaged - the basis of reasoning across all domains of cognition.

\footnotetext{We note that there are many other theoretical accounts of cognitive maps which do not address this issue of representation learning \cite{Radulescu2021, Sanders2020a, Stoianov2020}. While these models have provided mechanistic insights, they are not discussed in detail here.}

\section*{The cognitive mapping problem}

Cognitive maps organise knowledge to afford flexible behaviour \cite{Tolman1948, OKeefe1978, Behrens2018, Niv2019}. Affording \textit{behaviour} means that the cognitive map must contain information relevant to downstream behavioural tasks. Affording \textit{flexibility} means the map must 1) afford new behaviours in the face of new challenges, and 2) be built as fast as possible, ideally immediately, for any new world - a concept known as \textit{systematic generalisation} in the machine learning literature \cite{Bahdanau2018}. The aim for cognitive maps, then, is to learn as much as possible ahead of time, so online learning and computations are minimised - in essence, \textit{learning so that you do not have to learn}. In order to achieve this, there are some requirements and desiderata for the neural representations of the cognitive map. These computational considerations have led to models of the system which have had many recent successes in predicting neuronal representations. Here we describe these computational considerations and explain (in linked boxes) the models relevant to each. We aim to provide a clear conceptual understanding of the interlinked ideas. .

\subsection*{Reinforcement learning and planning}

To afford successful behaviour, cognitive maps must represent \textit{state} (a particular configuration of the world). Knowing when to turn right or left while driving requires an understanding of the orientation of the steering wheel, how far pedals are pushed, the curvature of roads, content of road signs, the location of other vehicles, etc. - turning left when the road bends right can be problematic, after all. Reinforcement learning (RL \cite{Sutton2017}) is a formalism of this concept: actions are taken based on the current world state (for instance, turning right when the road bends right). Representing the entire world state (in this analogy, not only the steering wheel but also the positions of planets, etc.) is often infeasible, as it can contain information along countless dimensions - and often dimensions irrelevant for solving the current task. Not only is this problematic for representational capacity, but it also impedes the efficiency of learning (it should not be necessary to re-learn how to drive when moving from a red to a black car). This effect, known as the `curse of dimensionality' \cite{Bellman1957}, can be mitigated by an appropriate state \textit{abstraction} (for instance, ignoring colour in the case of cars). Learning, or attending to, the appropriate abstraction is a central issue of the cognitive mapping problem \cite{Gershman2010, Wilson2014a, Radulescu2021}.

Classic (or \textit{model free}) RL learns the value of states, or which actions are good in which states, and therefore requires no knowledge of how states relate to each other. While this is provably optimal in the long term \cite{Watkins1992}, value-based learning is often inflexible and slow to learn \cite{Sutton2017}. Knowing relationships between states - that is, knowing the state-space structure - however, lets you play different games. Now you can flexibly \textit{plan} routes between any start and any goal state. For instance, taking never-before-experiences routes home \cite{Tolman1946} (that is, even if those states have no values attached to them) should one's normal route home be blocked. Unfortunately, traditional planning mechanisms (such as tree-search) are computationally costly, but alternatives do exist \cite{Silver2018a, Botvinick2012, Bush2015}. More broadly, with a clever representation of the state-space (see next section), the cost of planning can be reduced, or even completely avoided. This is a powerful way to formalise the central goal of cognitive maps: building maps that help solve problems in representation, not by exhaustive computation.

\subsection*{Space as a state-space}

To understand what this means, let us consider the state-space of physical space. Here, rather than representing sensory configurations, the state-space comprises physical locations like a literal map. This abstraction alone clearly profoundly helps the spatial planning problem. However, location can be represented in a variety of ways; for example, an unique identifier for each physical location, or (x,y)-coordinates. The choice of which representation to use has major consequences. For example, to find the shortest route: In the former, you need search through a series of neighbours. In the latter, you can compute a vector by subtracting start from end representations. Consider adding a new location. In the former, a new identifier (cell!) is required along with new relationships (synapses!) to neighbouring identifiers. In the latter, nothing new is required - since (x,y) naturally extends to new locations. These two representation types are analogous to place and grid cells in the hippocampal formation. Individual place cells code for unique locations, and thus \textit{new} place cells are required for new locations, while grid cells afford vector calculations \cite{Bush2015, Stemmler2015}) and naturally extend to new locations (albeit periodically). By a clever choice of \textit{representation}, grid cells prevent the need for \textit{computation}.

\subsection*{Non-spatial state-spaces}

While it is easy to intuit good state-spaces in physical space, this problem becomes less clear in non-space. One promising approach, derived from RL \cite{Foster2000, Gustafson2011}, is to cast spatial learning as understanding relationships on a graph (Figure \ref{fig:intro}b). In space, nodes of a graph define physical locations, and so edges between nodes exist if two locations are directly connected. Graphs, unlike literal maps, need not consider as-the-crow-flies distances, since roads and airplanes render distant locations more connected. Instead, they consider non-Euclidean distances based on the notion of `connectedness'. This is a re-conceptualisation of a map in terms of its connections (topology), as opposed to distances (geometry). Significantly, graphs also formalise non-spatial problems (Figure \ref{fig:intro}b-c). Family trees, social networks, atoms in molecules, and many other problems, all consist of relationships between entities and can be represented with graphs. Nodes in the graph no longer represent physical locations, but instead non-spatial locations - for instance, Alice is Bob's grandparent in a family tree. See Box \ref{box:sr} for graph-based state-space models.

Graphs define state-spaces and so afford value-based RL. They also afford planning; starting with Bob (characterised by a vector \( \vv \) - all zeros except a \( 1 \) at the Bob node element), and multiplying \( \vv \) by \( \mT \) (\( \mT \vv \); \( \mT \) is the transition matrix, where \( \emT_{ij} \) is the transition probability from state \( j \) to \( i \)), gives a distribution over future states after one step. Similarly, multiplying again by \( \mT \) (\( \mT^2 \vv \)) gives the distribution after two steps. Repeating this process until a non-zero entry appears in the Alice node provides the shortest-path route between Bob and Alice (assuming you have the right transition matrix) - which is two since Alice is Bob's grandparent. Naturally, exactly the same tree search process works between two locations in space.

\begin{boxy}[title={Box 1: RL state-spaces, graphs, and graph representations}, label=box:sr]

\textit{The problem of building graphs for cognitive maps is the same problem as building state-spaces in reinforcement learning. Crucially, the state-space in RL is tightly linked to behaviour (through rewards, values and policies). However, once the state space is defined there is a further choice of how each state is actually \textbf{represented}. Clever choice of representation can reduce online value/policy computations. This has allowed normative mathematical theories to predict neural representations.}

Reinforcement learning is concerned with taking appropriate actions at specific states (\( s \)) to maximise the expected (discounted by \( \gamma \)) sum of future rewards \( v(s) = \mathbb{E} \left[ r(s) + \gamma r(s') + \gamma^2 r(s'') \cdots \right] \), where \( s' \) and \( s'' \) are states following \( s \). Bellman \cite{Bellman1957} realised that this is a recursive equation, since the right-hand side contains the left-hand side but one step in the future: \( v(s) = r(s) + \gamma \sum_t P(s' \mid s, \pi) v(s') \), where \( P(s_{t+1} \mid s_{t}, \pi) \) is the transition probabilities between states under a policy \( \pi \). In essence, Bellman's equation says the value of the current state is the reward at that state plus the average value of states you can transition to. If you can \textit{assign credit} to each state (like these equations do), then taking good actions is easy: just go to the neighbouring state with the highest value \( v(s') \).

RL state-spaces define graphs with transition matrix elements \( \emT_{ij} = P(s_{j} \mid s_{i}, \pi) \). One graph representation, the successor representation \cite{Dayan1993} (SR), is particularly relevant to cognitive maps \cite{Gustafson2011, Stachenfeld2017}. The SR is a (discounted) sum of \( n \)-step transition matrices - \( \mS = \sum_n \gamma^n \mT^n \). Elements of this matrix, \( \emS_{ij} \), describe connectedness via all possible paths between two locations. Critically, if we represent connections between states in the world in terms of the SR-distance, then computing value is easy, since the SR is one half of the value computation \cite{Dayan1993} (\( \vv = \mS \vr \) where \( \vv \) and \( \vr \) are vectors whose elements are values and reward at each state).

Stachenfeld and colleagues \cite{Stachenfeld2017} noticed that the rows of \( \mS \) look like hippocampal place cells, and the eigenvectors of \( \mS \) resemble entorhinal grid cells (Figure \ref{fig:sr}; similar to work demonstrating that the eigenvectors of place cell correlation matrices resemble grid cells \cite{Dordek2016}). Notably, SR makes many predictions about how both grid and place cells behave in different environments and tasks \cite{Mehta2000, Derdikman2009, Krupic2012, Stachenfeld2017}. Critically, it also makes predictions of representations in non-spatial tasks \cite{Garvert2017, Schapiro2017, Stachenfeld2017}. Because it derives from a theory of learning, it can also account for behavioural phenomena that are otherwise hard to explain \cite{Momennejad2017}.

One prominent issue with SR, however, is its policy-dependence \cite{Momennejad2020}. This means that when rewards move - or, worse, when obstacles appear - value calculations using SR are no longer optimal \cite{Momennejad2020}. A recent model addresses this problem \cite{Piray2020}, using linear RL \cite{Todorov2007}. This model builds a representation for default behaviours that can be linearly updated when rewards change to approximate the new optimal policy. The required default representation (DR) resembles the SR, and can therefore be computed from grids cells. The model further provides a novel account of how to build world representations \textit{compositionally} out of component cells representations (e.g. how grid and border cells interact to represent the insertion of a barrier) \cite{Mark2020}. We return to this important issue in box \ref{box:tem_smp} and related text.

\begin{center}
\captionsetup{type=figure}
\colorbox{white}{\includegraphics[width=0.7\textwidth]{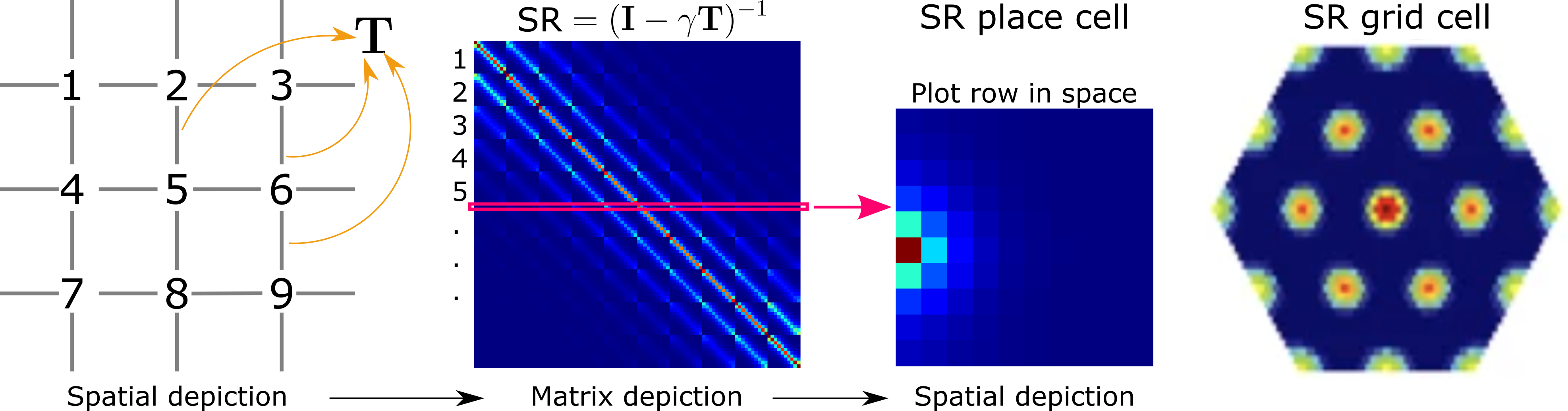}}
\captionof{figure}{\textbf{Left:} Graphs can be succinctly represented via their transition matrix, \( \mT \). \textbf{Centre:} From this, the SR matrix can be calculated \cite{Dayan1993}. \textbf{Right:} Rows of this SR matrix resemble hippocampal place cells, and its eigenvectors resemble periodic entorhinal representations such as grid cells \cite{Stachenfeld2017} (note this eigenvector is from a hexagonal, not a square, world).}\label{fig:sr}
\end{center}

\end{boxy}

\subsection*{Latent states and sequence learning}

Graphs are a flexible tool for representing problems, but how do we know which graphs to build? What should be the nodes in the graph or, equivalently, the states in the RL problem (Figure \ref{fig:intro}f)? A sensory observation alone cannot define a state, since two identical observations can have very different consequences; crossing a road implies looking right in the UK, but left in Germany. Formally, our world is not `fully observable'; instead, we face `partially observable' problems and must infer \textit{latent state} \cite{Gershman2010, Niv2019} representations that disambiguate UK and German roads. 

While individual observations are not enough to infer latent state representations, \textit{sequences} of observations are, since all identical observations do not have identical surroundings (remembering you just read a newspaper in German is enough to tell you to look left when crossing a road (Figure \ref{fig:intro}g)). See Box \ref{box:cscg} for clone-structured cognitive graph (CSCG) - a model that infers latent states from sensory sequences.

Neural representations in the hippocampal formation disambiguate states using latent representations \cite{Dusek1997, Wood2000, Frank2000, Komorowski2009, Carpenter2015, Grieves2016, Sun2020, Nieh2021}. For example, rodent grid cells will initially code identically for two identical boxes. However, after realising that the boxes are connected by a corridor, the grid representation changes to become consistent with the \textit{global} two-box-and-corridor-space \cite{Carpenter2015} (Figure \ref{fig:intro}d). These are latent state representations that disambiguate sensory aliased boxes due to their different futures. Physical location can also be aliased; in spatial alternation tasks \cite{Frank2000, Wood2000} (Figure \ref{fig:intro}e), the same physical position (in this case, the central stem of the maze) predicts different futures depending on animal's left/right choice on the previous trial. Splitter cells \cite{Frank2000, Wood2000}, place cells \cite{Komorowski2009}, grid cells \cite{Carpenter2015}, lap cells \cite{Sun2020}, and others, are all cellular examples of the cognitive map disambiguating the world into \textit{latent} states.

\begin{boxy}[title={Box 2: Building latent state representations from sequences}, label=box:cscg]

\textit{State-spaces must be inferred from observations. Because the sensory world is \textbf{aliased} - the same observation can occur more than once - states cannot be inferred from sensory appearance alone. Instead, \textbf{sequences} of observations uniquely identify states since two states with the same sensory observation will have different futures. States inferred via sequences are known as \textbf{latent} states, and building a latent state-space map can be used to afford different behaviours in sensorially identical situations.
}

The clone-structured cognitive graph (CSCG) model \cite{George2021} is an elegant approach for building de-aliased state-spaces. Here, hippocampus contains multiple `clone' cells for each sensory observation \cite{George2021, Cormack1987}. Now, one hippocampal `frog' clone cell responds to a frog in one location, and another responds if a frog appears elsewhere (Figure \ref{fig:cscg}). The model uses Bayes to 1) infer which hippocampal clone cells should be active for each sensory observation and 2) learn an appropriate set of transition weights between clone cells. These transition weights are analogous to the transition matrix for graphs, but critically the state-space is \textit{learned}, rather than provided by the modeller. 

Many hippocampal findings can be understood in terms of representing latent states, from basic phenomena, such as place cells, through to complex representations which vary as a function of animal behaviour. These predictions are in the main common between CSCG and more complicated models that follow, and we show a number of these in detail in figure \ref{fig:latent}. A critical difference between CSCG and the following models, is that CSCG infers the whole latent space within the hippocampus (as opposed to the cortical input to hippocampus). This enables learning rules to be local, biologically plausible, and fast. By contrast, CSCG has to learn each map \textit{de novo} and cannot benefit from having learnt similar maps before. It is exciting to think how these benefits may be combined (see section `Complementary maps in hippocampus and cortex', Figure \ref{fig:integrations}b).

CSCG is easily expressed in mathematics, and is closely related to hidden Markov models. From a sequence of sensory observations \( \sX = \{ \vx_1, \vx_2, \vx_3, \cdots, \vx_T \} \) and actions \( \sA = \{ \va_1, \va_2, \va_3, \cdots, \va_T \} \), we wish to infer \textit{discrete} latent states \( \sZ = \{ \vz_1, \vz_2, \vz_3, \cdots, \vz_T \} \). Now, the same sensory observation, \( \vx \), can be linked to different latent states (clones) \( \vz \), via an `emission' distribution \( p(\vx \mid \vz) \), naturally accounting for the aliasing problem. Along with predicting sensory observations, CSCG latent states predict future latent states and actions \( p(\vz_t, \va_t \mid \vz_{t-1}) \). Modelling the full sequence of observation is then:
\[ 
p(\sX, \sZ, \sA) = p(\vz_0) \prod_t p(\vx_t \mid \vz_t) p(\vz_t, \va_t \mid \vz_{t-1})
\label{eq:clones_prob}
\]
Here, each element of \( \vz \), \( \evz_i \), is a `clone' of a sensory observation (Figure \ref{fig:cscg}, note we use \(t\) for vectors in time and \(i\) for indexing elements of each vector). Concretely, if there are 4 possible sensory observations, and 5 clones for each observations, there will be 20 elements to \( \vz \). The probability of observing a `frog' given a `frog clone' is defined as 1, but 0 given a `snail clone' - \(p(\vx \mid \evz_i \in C(\vx)) = 1 \) whereas \(p(\vx \mid \evz_i \notin C(\vx)) = 0 \) if \( C(\vx) \) are the clones of \( \vx \). CSCG marginalises over \( \vz \) and uses the expectation-maximisation algorithm to train the model \cite{Dempster1977} - that is, learn an appropriate set of transition probabilities \( p(\vz_t, \va_t \mid \vz_{t-1}) \) and infer \( \vz_t \). Once trained, this model can be used for planning by inferring a sequence of actions and observations conditioned on a start and end clone.

\begin{center}
\captionsetup{type=figure}
\colorbox{white}{\includegraphics[width=0.65\textwidth]{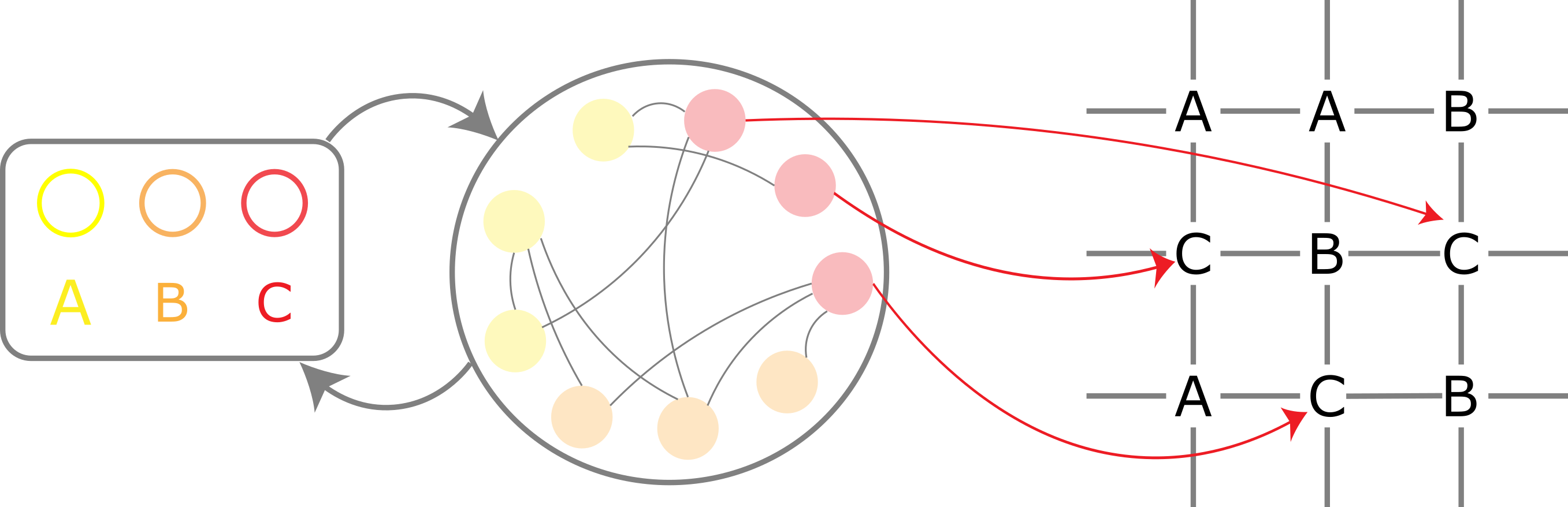}}
\captionof{figure}{CSCG \cite{George2021} addresses the problem of sensory-aliasing using multiple hippocampal clones for each sensory observation, then inferring which clone should be active at each location (as well as learning a clone transition matrix).} \label{fig:cscg}
\end{center}

\end{boxy}

\subsection*{Path integration and compression}

Inferring latent states is really a problem of understanding where you are in an abstract space. In the two-room task \cite{Carpenter2015}, the global grid code which forms uniquely identifies physical locations. For the spatial alternation tasks, `splitter' cells identify simultaneous location in physical space and location in left/right trials. Working out where you are in physical space is easy - accumulate self movement vectors (e.g. \texttt{North}, \texttt{South}, \texttt{East}, and \texttt{West} from head direction cells \cite{Taube1990}) to update your relative position: this is path integration \cite{Darwin1873} (Figure \ref{fig:path_int}a). Ants, rodents, birds, and humans all path integrate \cite{Mittelstaedt1980, Etienne2004, Loomis1993}, with mammalian path integration crucially dependent on the hippocampal formation \cite{Maaswinkel1999}. Entorhinal grid cells are considered an attractive substrate for path integration of two-dimensional spaces since 1) periodic representations extend to all, even unseen, space; 2) the periodicity of grid codes is inherently error-correcting \cite{Sreenivasan2011a}; 3) grid cells represent location with the same precision but far fewer cells than place cells \cite{Mathis2012}; 4) grid cells are experimentally driven more from path integration signals that place cells \cite{Chen2019}. See Box \ref{box:path_int} for path integrating models.

To work out where you are in graphs and non-space requires a modification of path integration. Rather than accumulating self-movement vectors, accumulate abstract movement vectors instead, (\texttt{Parent}, \texttt{Child}, \texttt{Sibling}, \texttt{Aunt}, \texttt{Nephew}, etc. for family tree graphs). Just like (x,y)-coordinates versus unique identifying representations for space, path integrating representations on graphs offer a benefit compared to representing every individual connection between nodes: adding a new node (Chloe is the \texttt{Sibling} of Bob) immediately implies all other connections (Chloe is the \texttt{Grandchild} of Alice) without needing to observe that relationship explicitly. This is because, just like (x,y)-coordinates, path integration treats all nodes equally and relationships are structured (\texttt{Sibling} + \texttt{Grandparent} = \texttt{Grandparent}). As such, only the few rules of path integration need to be known, not every possible relationship; path integration is a \textit{compressed} representation. Not all graphs, however, can be path-integrated, since consistent actions do not always exist across graphs (for instance, social networks merely describe generic relationships). 

\begin{boxy}[title={Box 3: Path integrating state-spaces}, label=box:path_int]

\textit{Path integration offers a powerful way to build latent state spaces. It builds maps that embed knowledge of the structure of the space (in physical space, \texttt{North} + \texttt{East} + \texttt{South} + \texttt{West} = 0). This means that path integration maps are: 1) inherently latent (and abstract), since they follow rules, not sensory observations and 2) allow relational knowledge to be transferred to any situations where the same rules apply. Notably, although path integration is not limited to space, not all graphs can be path-integrated.}

Path integrating models utilise a particular type of recurrent neural network (RNN) known as continuous attractor neural networks (CANNs \cite{Zhang1996}; Figure \ref{fig:path_int}b), where neurons are recurrently connected via weights, \( \mW \), and receive velocity input, \( \va \).The neural dynamics are given by:
\[ 
\tau \frac{d \vg}{dt} = -\vg + f \left ( \mW \vg + \mB \va \right)
\label{eq:cann}
\]

Where \( \tau \) is the time constant of neuronal response and \( f \) a non-linear activation function. 
\footnotetext{An alternative, but less biologically plausible, equation is \( \tau \frac{d \vg}{dt} = -\vg + f \left( \mW_{\va} \vg \right) \), where the recurrent matrix \( \mW_{\va} \) depends on the movement velocity.}
With an appropriate set of weights, CANNs path integrate, with different cell classes (head-direction cells \cite{Skaggs1995, Zhang1996}, place cells \cite{Samsonovich1997a, Tsodyks1999a}, grid cells \cite{Burak2009}; Figure \ref{fig:path_int}d) modelled with different weights. Remarkably, CANNs really exist in nature; ring attractors \cite{Ben-Yishai1995}, both in connections and anatomy, are found in flies \cite{Kim2017}, and attractor manifolds are found in rodents \cite{Yoon2013, Gardner2021}.

Other path integrating models exist \cite{OKeefe1993, Burgess2009}. For example, velocity-coupled oscillators (VCOs) suggest path integration (along an axis) via interference between theta oscillations and velocity-dependent dendritic oscillations, with their phase difference indicating path integrated distance along an axis (this looks like a plane wave!). Here, grid cells are the sum of three such neurons with preferred axes at \( \frac{\pi}{3} \) relative angles. 

One major limitation of CANNs and VCOs, however, is that the weights of the recurrent weight matrix, \( \mW \), are carefully selected and not learned from sensory experience. However, it is easy enough to set up path integration as a learning problem via predicting observations \( \vx\): path integrate the latent state variable \( \vz \) and then predict observations \( \vx \) from the latent states:
\[
p(\sX, \sZ \mid \sA) = p(\vz_0) \prod_t p(\vx_t \mid \vz_t) p(\vz_t \mid \vz_{t-1}, \va_t)
\label{eq:path_int_prob}
\]
Where the path integrating part (\( p(\vz_t \mid \vz_{t-1}, \va_t)\)) is now replaced by a discrete-time version - i.e. \(\vz_t = f \left ( \mW \vz_{t-1} + \mB \va \right) + noise \). In fact, several models use a deterministic RNN (i.e. set the noise term to 0). These models successfully learn to path integrate when tasked with predicting ground truth spatial representations - i.e. \( \vx \) is either place cells \cite{Banino2018}, or (x,y) coordinates \cite{Cueva2018}. Neural units in both models form periodic representations (Figure \ref{fig:path_int}e-f), but these are often amorphous, four-fold symmetric grids. An elegant analytic result \cite{Sorscher2019}, however, demonstrated that the 4- to 6-fold symmetry transition is governed by a single property: a third order regularisation term of grid cells. Indeed, this is easily implemented by the biological constraint of ensuring neural activity is positive \cite{Dordek2016, Sorscher2019}.

\begin{center}
\captionsetup{type=figure}
\colorbox{white}{\includegraphics[width=0.95\textwidth]{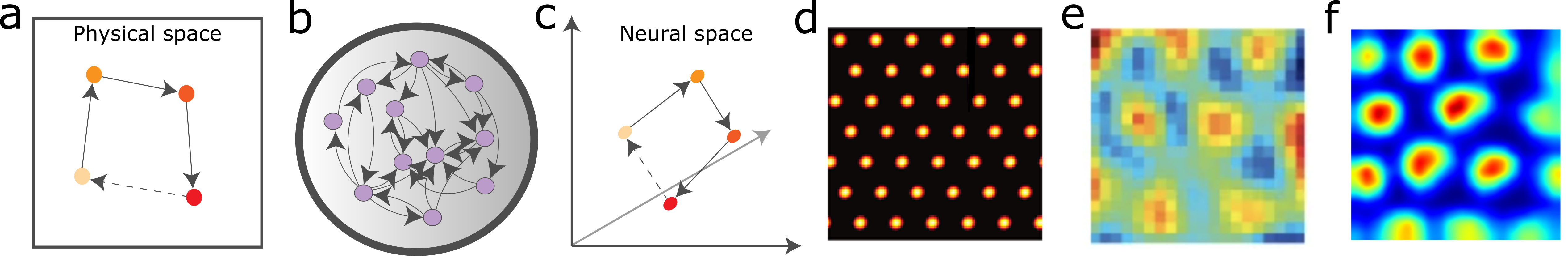}}
\captionof{figure}{\textbf{(a)} Path integration is the summation of self-motion vectors to permit self-localisation. \textbf{(b)} CANNs are recurrently connected neurons which \textbf{(c)} path integrate, and \textbf{(d)} exhibit representations mirroring grid cells \cite{Burak2009}. \textbf{(e)} In LSTMs \cite{Banino2018} and \textbf{(f)} other RNNs \cite{Sorscher2019}, grid-like representations are learned when trained to predict ground-truth spatial representations.} \label{fig:path_int}
\end{center}

\end{boxy}

\subsection*{Generalisation}

Generalisation, or the transfer of knowledge from one situation to another, is the substrate of the profound behavioural flexibility exhibited by animals. Without it, new situations could not be understood in the context of existing knowledge, and so prior learned behaviours could not be leveraged in new situations. Generalisation in the sensory domain means realising that a Pekingese and a Rottweiler are both types of dog. In the structural domain, however, deep and powerful inferences can be made: doors often lead to new rooms; addition works for 100s as it does for 10s; the same path integration rules apply across different spaces. These pieces of knowledge have profound effects on behaviour.

Generalising with graphs, however, is hard as they require perfect alignment - if the bottom left of the new environment is prescribed the old environment's top right representation, then the rest of the environment cannot be represented, since the graph ends. Perfect alignment (\textit{graph matching}, analogous to structural mapping \cite{Gentner1983}), however, is NP-hard and thus impractical in most situations. Generalising with periodic path integration representations, on the other hand, is easy since all positions are treated equally - representations corresponding to the bottom right in one environment could equally represent the middle of another environment. Furthermore since path integration maps are latent (thus abstract) by nature, they chart the relational structure of one family just as well as for another. This is generalisation of \textit{relational} knowledge.

The hippocampal formation is critical for generalisation, along with memory, and some forms of imagination \cite{Scoville1957, Cohen1980, Hassabis2007}. Hippocampal representations, however, do not generalise; neighbouring place fields are not necessarily neighbours in other environments, a phenomenon known as \textit{remapping} \cite{Anderson2003, Bostock1991, Muller1987} (Figure \ref{fig:intro}c). Entorhinal representations, on the other hand, do generalise; neighbouring grid cells (within-module) are still neighbouring grid cells in other environments - that is, the map is shifted and/or rotated (grid cell realignment \cite{Fyhn2007, Yoon2013}); representations corresponding to the bottom right in one environment now equally represent the middle of another environment. Spatial generalisation, at least, seems to exist in entorhinal cortex and is consistent with path integration.

Learning to generalise is often also a sequence-learning problem, but with sequences from \textit{many different environments} (Figure \ref{fig:intro}g). When hearing about new family, after observing Daniel is Emily's parent, and Fran is Daniel's sibling, it is \textit{only} possible to predict Fran's niece (Emily) with \textit{a priori} learned relational knowledge: \texttt{Parent} + \texttt{Sibling} + \texttt{Niece} = 0. 

To actually make sensory predictions, however, you need to know more than just abstract knowledge. You need to know how it interacts with real world representations (the abstract location in a family tree interacts (corresponds) with Emily for one family, and Chris for another; Figure \ref{fig:tem_smp}a). One influential proposal is that hippocampal cells reflect this interaction, with abstract knowledge from medial entorhinal cortex (MEC) and sensory knowledge from lateral entorhinal cortex (LEC) combined rapidly (`fast-mapped' \cite{Carey1978}) in hippocampus \cite{Manns2006, Behrens2018, Whittington2020}. This bridges the abstract-to-real divide and permits generalisation, since the \textit{same} abstract map can be reused across different sensory (LEC) environments and contexts. See Box \ref{box:tem_smp} for models that \textit{generalise}.

\subsection*{Composition}

Generalisation is more than transferring a single map to a new environment. Often only sub-components, or combinations of sub-components, need to be generalised. For example, understanding differently shaped rooms can be broken down into two components - an underlying 2D space and walls that can be placed anywhere. Should the cognitive map represent such common structural elements across tasks, then these elements can be \textit{composed} to understand any given task configuration \cite{Piray2020, Mark2020, Whittington2020}. To encourage arbitrary composition, different structural elements (bases) should be represented independently (factorised) from one another \cite{Behrens2018}. Understanding a task then becomes a structural inference problem - finding the appropriate bases to represent the current task \cite{Kemp2008, Lake2015}; this is a form of structural analogy \cite{Battleday2020}. The value of such an approach has already been demonstrated in cognitive models which formalise composition in other domains of cognition, such as language and algebra \cite{Lake2015, Ellis2020}. 

The cognitive map of the hippocampal formation contains many such basis representations (Figure \ref{fig:intro}a,c). Object-vector cells \cite{Hoydal2019} (OVCs), border-vector cells \cite{Hartley2000, Becker2001, Barry2006, Solstad2008, Lever2009} (BVCs), reward cells \cite{Gauthier2018}, and goal-direction cells \cite{Sarel2017} (GVCs) , are all examples of \textit{local bases} - representations that encode any object/border/goal, irrespective of where it is. Grid cells, by contrast, are examples of \textit{global bases}, as they describe information equally across all space.

\begin{figure}
\centering
\includegraphics[width=0.93\linewidth]{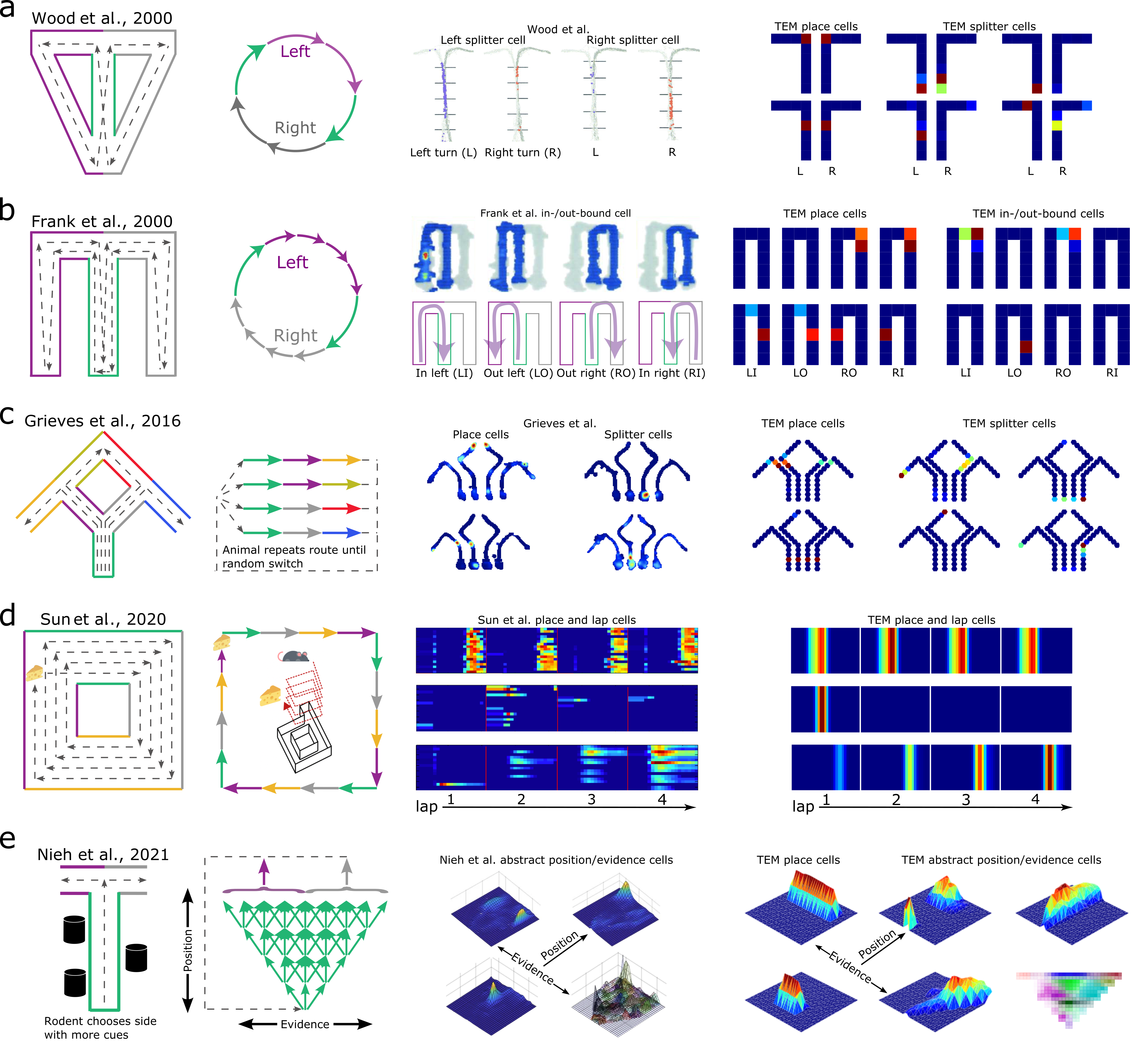}
\caption{\textbf{Representing latent states.} Many apparently different neural phenomena are captured with a unifying computational principle - building state-spaces that can accurately predict different futures (latent states) as fast as possible (generalisation). \textbf{(a-e)} For each row, left/center-left are the task and its latent state-space (with colours denoting sensory experience), while center-right/right are real/TEM neural representations. \textbf{(a)} In a T-maze task \cite{Wood2000}, where animals alternate left/right turns, the state-space is described by a `big-loop' latent space, since the central trunk predicts different futures depending on a previous left/right turn. Hippocampal cells represent both space (place cells) and the `big-loop' (splitter cells). Splitter cells are trajectory-dependent, firing at the \textit{same} spatial location (central trunk) \textit{differently} depending on the prospective future (left/right). TEM learns both spatial (place) and non-spatial (splitter) cells when trained on this task; splitter cells to represent latent state in the `big-loop' and place cells to represent physical location, and thus facilitate spatial generalisation. \textbf{(b-c)} More complicated spatial alternation tasks \cite{Frank2000, Grieves2016} are also described with `big-loop' latent state-spaces. Both real and TEM hippocampal representations contain spatial (place) and non-spatial (trajectory-dependent) cell representations. \textbf{(d)} Performing 4 laps to receive a reward is a non-spatial task \cite{Sun2020}. It is also described as a `big-loop' latent state-space. Rodent hippocampus, and TEM, represent both space (place cells; top) and non-space (lap-specific cells; middle/bottom). \textbf{(e)} A T-maze task where rodents choose left/right depending on sensory evidence (as the animal moves along the central trunk) has a latent state-space spanned by position and evidence. Hippocampal cells, and TEM learned hippocampal representations, map this position-evidence latent space i.e. not just spatial location (the bottom right panel is a collection of many different cells representations). We note that CSCG would learn the non-spatial cells (e.g. splitter cells) but not the spatial cells, as it is not a model of generalisation. Code for simulations will be made available on publication.}
\label{fig:latent}
\end{figure}

\begin{boxy}[title={Box 4: Generalising with memories}, label=box:tem_smp]

\textit{We have seen models that build latent state representations, and models that path integrate. If these principles could be combined, we could build a powerful system that learns arbitrary latent states from sensory observations (like CSCG \cite{George2021}) but additionally \textbf{generalises} these representations (like path integration models \cite{Burak2009,Banino2018}) and composes them arbitrarily. For \textbf{abstract} representations to be reused (generalised) in different \textbf{sensory} environments, the \textbf{same} abstract locations must be `linked' to \textbf{different} sensory observations. Hippocampal memories offer the ideal substrate for this link - they can rapidly tie sensory observations to specific locations.}

Hippocampal models of generalisation (the Tolman-Eichenbaum machine, TEM \cite{Whittington2020}, and the spatial memory pipeline, SMP \cite{Uria2020a}) are tasked with predicting, as fast as possible, sensory observations in novel, but structurally similar, environments (for example, multiple different families or 2D worlds; Figure \ref{fig:intro}g). Both models consist of two key components: 1) An abstract path-integration module that is reusable across environments. 2) A \textit{relational memory} \cite{Cohen1980} module that, like an address book, links abstract location representations with sensory representations (Figure \ref{fig:tem_smp}a). These links change from world to world, allowing the same abstractions to apply to multiple worlds. 

Recall the probabilistic interpretation of path integration:
\[
p(\sX, \sZ \mid \sA) = p(\vz_0) \prod_t p(\vx_t \mid \vz_t) p(\vz_t \mid \vz_{t-1}, \va_t) 
\label{eq:path_int_prob2}
\]
Previously, \(p(\vz_t \mid \vz_{t-1}, \va_t)\) was fixed and so each abstract location \(\vz \) could only predict a single sensory observation \(\vx \). If, instead, we had an address book of relational memories \( \mM \), we could remember what is where in \textit{each} environment. To predict upcoming sensory observations, all that is required is to imagine a transition in abstract representation (\( \vz\), via path integration), then retrieve the memory at that location (`what' did I see the last time I was `here'). Sensory prediction is now a combination of path integration and memory retrieval. 
But what space are we path integrating in, and how does it get built? Previously the weights, \( \mW\), in the path integrator (\( p(\vz_t \mid \vz_{t-1}, \va_t) \), where \( \vz_t = f \left ( \mW \vz_{t-1} + \mB \va \right) + noise ) \)) were built from predicting (x,y) coordinates or place cells (i.e. spatially curated representations). Now, we can predict actual sensory observations. This is more powerful. When sensory objects are arranged in space, the same spatial path integration as previous models will be learned, but when the sensory world has more complex dependencies, these will also be learnt. If the best way to predict the sensory future is to learn a complex map of latent states, then these models will learn to path integrate in this latent space (Figure \ref{fig:latent}).

While TEM and SMP are conceptually the same model, they have different implementations. Two critical ones are (1) that TEM is supplied with allocentric actions and object representations, but SMP must infer them from egocentric input and pixels; (2) that SMP implements memory with a memory network from machine learning \cite{Pritzel2017a}, while TEM uses more biologically realistic Hebbian learning \cite{Hebb1949} and Hopfield networks \cite{Hopfield1982}. This biological constraint means that the link between abstract and sensory world must take place in neuronal units. That is, the same hippocampal neurons must know \textit{both} the abstract location and the sensory prediction. This type of \textit{conjunctive} representation is commonly observed real in hippocampal neurons \cite{Komorowski2009, McKenzie2014a}.

TEM and SMP are deep artificial neural networks which learn to generalise structural knowledge and recapitulate a host of known representations of the hippocampal cognitive map in doing so (Figure \ref{fig:tem_smp}c-d). Since SMP works from ego-centric inputs, it generates cells involved in the ego- to allo-centric coordinate transformation \cite{Becker2001}. Additionally, TEM learns \textit{compositional} entorhinal representations in spatial and non-spatial tasks, and can solve classical relational memory tasks that are crucially dependent on the hippocampal formation, such as transitive inference \cite{Bunsey1996, Dusek1997}.

\begin{center}
\captionsetup{type=figure}
\colorbox{white}{\includegraphics[width=0.95\textwidth]{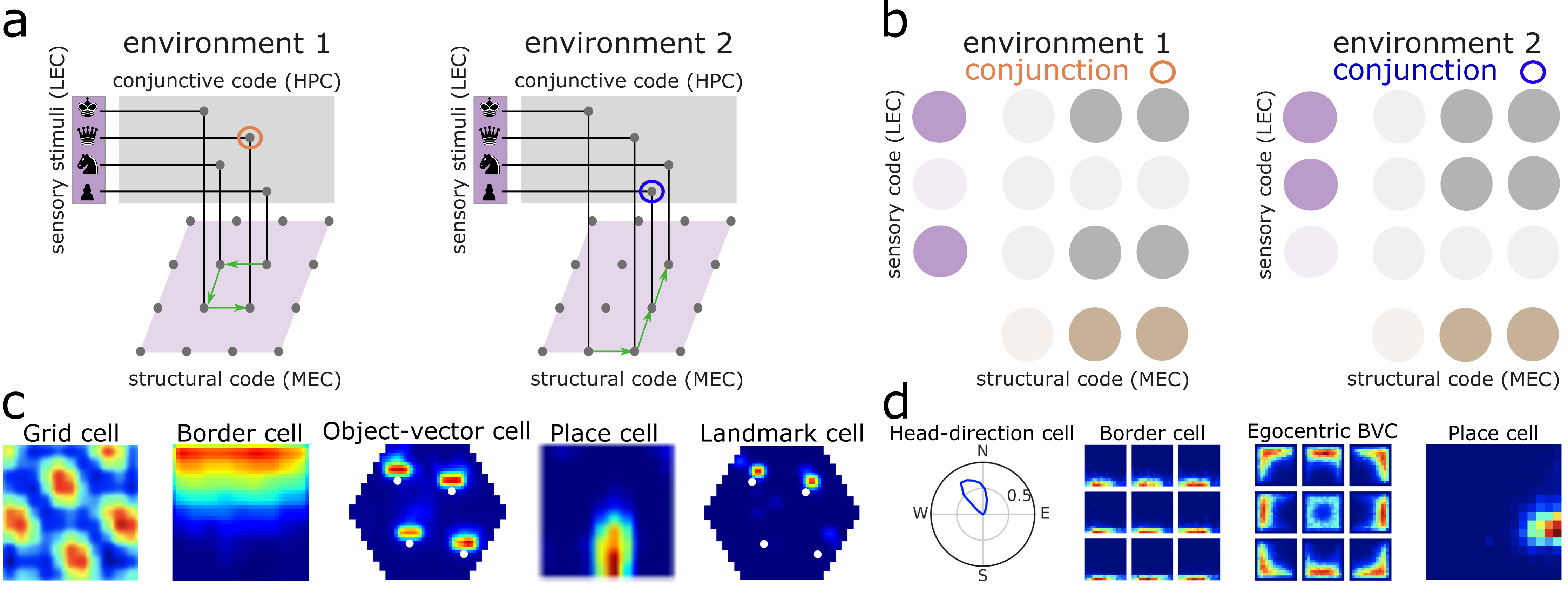}}
\captionof{figure}{\textbf{(a)} Schematic of models (adapted from Sanders \textit{et al.}\cite{Sanders2020}), where the \textit{same} cortical representations (LEC and MEC) are reused in different environments, facilitated by \textit{different} hippocampal combinations. \textbf{(b)} TEM conjunctive hippocampal cells receive input from particular MEC and LEC cells; hippocampal remapping affords generalisation. \textbf{(c)} TEM and \textbf{(d)} SMP recapitulate a host of empirically described cell representations \cite{Whittington2020, Uria2020a}.}\label{fig:tem_smp}
\end{center}

\end{boxy}

\section*{Novel interpretations, integrations, and predictions}

While the models above account for a variety of data, including hippocampal and entorhinal cellular representations from spatial and non-spatial tasks, they often do so in seemingly divergent ways, and there are many neural phenomena that remain perplexing. Here we consider how these ideas can be integrated in order to model and understand cognitive maps at a deeper level, and offer novel accounts of several neural phenomena through a formal lens.

\subsection*{Non-spatial hippocampal cells are latent state representations for generalisation}

Many non-spatial hippocampal representations have appeared over the last few decades \cite{Wood2000, Frank2000, Grieves2017, Sun2020, Nieh2021}. These cells appear to represent different tasks in different ways, but we can understand all these cell types with a single framework: representing latent state-spaces. We have argued latent state representations serve two purposes; firstly, separating states that have different futures and, secondly, affording generalisation, since the latent map can be used across multiple different environments. These arguments suggest two things: 1) Hippocampal and cortical representations do not map abstract spaces for the sake of it, but only to disambiguate states with different futures. 2) To generalise as fast as possible, every level of abstraction needs to be represented simultaneously; representing space in spatial tasks, non-space in non spatial tasks, and \textit{both} space and non-space in interacting spatial-non-spatial tasks.

As a didactic example, consider spatial alternation tasks \cite{Wood2000, Frank2000} where animals cycle \( left \to right \to left \to right \cdots \) at a choice point (Figure \ref{fig:latent}a). This task can be `un-rolled' into a `big-loop' state-space where the first half is going left and the second is going right. This is a latent state-space for the task; it de-aliases the common `trunk' section according to whether the animal is going to take a left or right turn. This `big-loop', however, ignores spatial knowledge - understanding the big-loop \textit{alone} does not let you \textit{know} you are back in the same place each time you return to the common section - to generalise spatial knowledge you additionally need a spatial representation. Hippocampal cells in this task indeed code for both space (place cells) and big-loop (splitter cells) \cite{Wood2000}.

This interpretation of non-spatial hippocampal representations (latent states for disambiguation and generalisation) can be modelled formally. Here we show many non-spatial tasks \cite{Wood2000, Frank2000, Grieves2017, Sun2020, Nieh2021} can be understood by these two principles alone. We use TEM, since it learns and generalises latent states at multiple levels of abstraction. We note that pure latent state models, such as CSCG, account for the separation of states with different futures (e.g. it learns splitter cells for the big-loop), but without extra assumptions it will not learn place cells at the same time, as it cannot profit from generalising the structure of space.

Firstly, training TEM on spatial alternation tasks \cite{Wood2000, Frank2000, Grieves2017} (Figure \ref{fig:latent}a-c), TEM recapitulates both splitter cells and place cells. Splitter cells for the `big-loop' and place cells for spatial generalisation. Secondly, when rodents are trained on a task where reward is received every 4 laps of a loop, hippocampus \cite{Sun2020} contains both spatial `place' cells that care about location in the lap, and non-spatial cells that additionally care about \textit{which} lap. This is exactly what TEM learns, too (Figure \ref{fig:latent}d) - lap cells for the `big-loop' and place cells for spatial generalisation. Lastly, when animals are trained to make left/right choices on a T-maze depending on the difference in number of sensory cues appearing on the left/right as the animal moves forwards in the central trunk (Figure \ref{fig:latent}e), hippocampal cells form an abstract map spanned by physical space and cue difference (termed `evidence'). TEM learns exactly this; physical space to predict when to make the choice, and cue difference to predict reward left or reward right.

\subsection*{Complementary maps in hippocampus and cortex} \label{cscg_tem}

The reviewed models mirror an age-old debate in the hippocampal literature - does hippocampus map space \cite{OKeefe1978}, or is its role one of memory \cite{Scoville1957, Eichenbaum2014}. TEM and SMP suggest memories since they adhere to hippocampal indexing theory \cite{Teyler2007}, where the hippocampal representations are simply an index that binds together cortical representations (structural (MEC) and sensory (LEC) representations in the case of TEM and SMP). In these models, hippocampus acts as a `memory map', forming memories at distinct locations in the map, but the notion of `map' is inherited, with all predictive capabilities occurring through cortex. These models embody an extreme version of Eichenbaum's view that the hippocampal role in navigation is principally one of relational memories \cite{Eichenbaum2014}. By contrast, in the state-space models (such as SR, DR, and CSCG), the hippocampus is an explicit map, where connections between hippocampal cells define the connections of the map. Regardless, both sets of models explain many hippocampal representations and phenomena.

\begin{figure}
\centering
\includegraphics[width=0.99\linewidth]{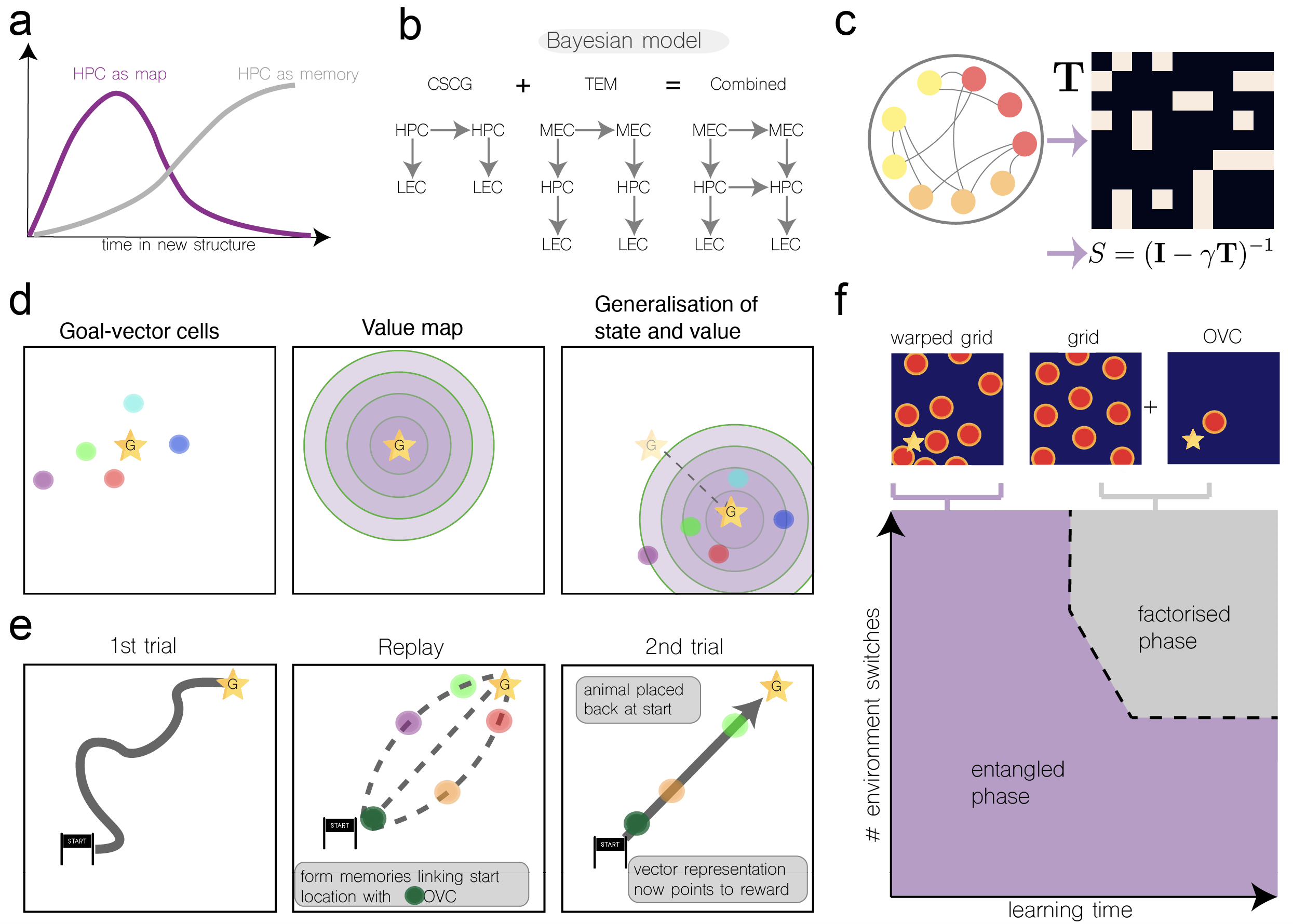}
\caption{\textbf{Integrating different cognitive map models and novel predictions.}
\textbf{(a-b)} The reviewed models suggest two roles for hippocampus: 1) A map - connections between hippocampal cells encode relationship between states - and 2) memories linking \textit{cortical} map representations. \textbf{(a)} We suggest hippocampus serves \textit{both} roles, but does so in different situations. In experiences where no prior cortical map is useful, hippocampal representations build a relational map; in familiar experiences where cortex has already learned how to structure (e.g. path integration) representations, hippocampus fulfils the role of memory. We suggest that with increasing experience, there is a transition from hippocampus as a map to memory, and this will be tied to the behavioural ability to generalise (via cortex). \textbf{(b)} TEM (HPC as memories) and CSCG (HPC as map) models can be easily integrated (since both are formalised probabilistically) into a model with a hippocampus that can form both maps and memories. \textbf{(c-e)} State-spaces for behaviour. \textbf{(c)} Learned latent state-spaces can be inputted into RL algorithms such as the successor representation. \textbf{(d)} On the other hand, compositional representations, such as GVCs, permit rapid generalisation of policy. Since these representations already generalise to novel goals in novel environments, all that is required is a pre-computed set of values (or policies) associated with the GVCs. The value map (or policy) is simply transferred along with the GVCs: \textit{credit assignment through generalisation}. \textbf{(e)} Replay might play a role in this mechanism. After encountering a goal, we want the goal-vector representations to exist across all of space, and especially any start locations. Replay trajectories provide an offline solution; path integrate (offline) GVCs and bind them (via memory) to important locations such as the start state. Thus, when re-entering the same environment, vector representations and the associated value map (or policy) already exist. This is replay as the offline building of maps for credit assignment through generalisation. \textbf{(f)} The aforementioned mechanisms rely on compositional (factorised) representations. Sometimes, however, brain representations are not compositional, but entangled \cite{Boccara2019}. Since compositional representations are beneficial for generalisation, we suggest animals have factorised or entangled representations depending on the pressure to generalise; regularly staying in the same task will encourage entangled representations, while regularly switching tasks will encourage factorised representations.}
\label{fig:integrations}
\end{figure}

The observation that the models of generalisation (TEM and SMP \cite{Whittington2020, Uria2020a}) treat hippocampus as memories alone, while models of single environments (SR, CSCG, DR \cite{Stachenfeld2017, Piray2020, George2021}) treat hippocampus as a map, is a distinction that offers a potential unification of the hippocampus role in mapping and memories: it is easier to learn how to generalise if each (latent) state-space is already built. More precisely, should all states of the world be appropriately separated, and relationships between states known, cortex can receive high-fidelity training signals (since predictions can be compared to a de-aliased state-space), thereby significantly reducing the burden of learning. This means entirely novel sets of relationships can be efficiently learned as follows: first, form temporary hippocampal maps; next, learn the statistical structure of these maps in cortex (Figure \ref{fig:integrations}a). This proposal follows complementary learning systems theory \cite{Marr1971, McClelland1995}, where cortex slowly learns the statistics of hippocampal episodes. We take note of an interesting model \cite{Evans2019} that, while not involving structural learning or generalisation, leverages two independent systems for self-localisation: relational maps in hippocampus and path integrating maps in entorhinal cortex.

This integrated approach is realisable within the existing models (Figure \ref{fig:integrations}). CSCG can disambiguate sensory states and rapidly learn hippocampal relational maps for \textit{single} tasks, while TEM (or SMP) learns path integrable abstractions that can be generalised between \textit{many} structurally similar tasks. Since both TEM and CSCG utilise multiple `clone' hippocampal cells for each sensory observation, it is particularly easy to combine these models. This would be formulated as a TEM-like model, but where hippocampus is predictive of future hippocampal states (Figure \ref{fig:integrations}b). Such an approach combines the best of both models - learning novel maps fast (CSCG), but also leveraging past knowledge to understand similarities between maps (TEM/SMP).

\subsection*{Cognitive maps and behaviour}

The models discussed here interact with behaviour in different ways (see Box \ref{box:eigenspace} for additional discussion on using eigenspaces for various behaviours). The models formulated in the RL framework (SR and DR \cite{Stachenfeld2017, Piray2020}) can be explicitly used for model-free and model-based learning, and so provide insight into the hippocampal role in constructing state-spaces for RL; constructing a \textit{predictive} cognitive map \cite{Stachenfeld2017, Gustafson2011} \footnotemark. Since these models only require well-separated state-space as input, any model that learns and builds state-spaces, such as CSCG, SMP, and TEM, could also act as input to these RL models. For example, the discrete state-space of CSCG (or TEM hippocampal cells) can be used for SR (Figure \ref{fig:integrations}c). 

\footnotetext{It remains unclear, however, whether hippocampal neurons could represent the vanishingly small differences in SR between (sometimes adjacent) states necessary for accurate reward-guided behaviour.}

The sequential models, such as CSCG, offer an alternative to tree search - `planning by inference' \cite{Botvinick2012, Friston2009}. This involves conditioning a probabilistic model on a start and goal state, then inferring a distribution over action sequences and their intermediary states. Since CSCG is a Bayesian model, it can naturally implement this procedure, thereby suggesting a hippocampal role in action inference. In principle, any probabilistic model with states and actions can be used to perform action inference \cite{Levine2018} (e.g. TEM), but the practicality of this differs across models and the details of their implementations.

For models that learn grid codes, vector-based planning can be used (for example, SMP and other neural network models implement vector-based navigation \cite{Bush2015, Banino2018, Uria2020a}). These models perform navigation and short-cutting behaviours reminiscent of biological agents, along with transferring policies from one environment to another (which is possible because these representations generalise).

\begin{boxy}[title={Box 5: Eigen-spaces},
label=box:eigenspace]

\textit{The observation that grid cells resemble eigenvectors of place cells (or of the spatial transition matrix) has led to interesting suggestions about mechanisms for planning and exploration.}

To plan the future, you need to look across multiple transitions. Eigenvectors simplify this problem because all multi-step transition matrices share the same Eigenvectors. We can see this by diagonalising the transition matrix \( \mT = \mV \Lambda \mV^T \), where \( \mV \) is a matrix with eigenvectors as columns, and \( \Lambda \) is a diagonal matrix of eigenvalues. A 1-step transition of a state vector \( \vs \) ( \( \mT \vs = \mV \Lambda \mV^T \vs \) ) uses the same eigenvectors as a 2-step transition (\( \mT^2 \vs = \mV \Lambda \mV^T \mV \Lambda^n \mV^T \vs = \mV \Lambda^2 \mV^T \vs \) since \( \mV^T \mV = \mI \)), and so on. The eigenvectors for \textit{local} and \textit{non-local} transitions are the same.

Intuitively, this means these eigenvectors can be used for exploration, planning, sampling in replay, or any other type of multi-step navigation. Different sampling patterns differ only in the eigenvalue matrix \( \Lambda \). When both eigenvalues and eigenvectors come from simple diffusion, navigation/ exploration strategies visit surrounding locations one by one, progressing further afield slowly just like diffusion. On the other hand, by making a clever alternate choice of eigenvalue matrix (using a bespoke diagonal matrix, \( \Upsilon \), rather than the diffusion eigenvalues, \( \Lambda \), but still using the same eigenvectors; \( \mV \Upsilon \mV^T \vs\) ), very different strategies emerge \cite{Mcnamee2021}, such as turbulence or super-diffusion (also known as L\'evy flights, which are known to be used by animals exploring environments for food \cite{Sims2014}), and can be seen in rodent hippocampal replay \cite{Pfeiffer2015}.

In fact, with another choice of weighting matrix, \( \Upsilon = \sum_n \gamma_n \Lambda^n \), you can exactly compute the SR under a diffusive policy, which is closely related to the distance between states. This is particularly interesting, as when you have distances planning is easy - just go the the neighbouring state with lowest distance from the goal. Importantly, this planning is by eigenvectors (grid cells) alone - in fact all you need for this `intuitive planning' \cite{Baram2018} are the start and goal grid codes.

So far we have been considering diffusive transition matrices, i.e. matrices without actions. However, by making transition matrices actions dependent (remember path integration has action dependent matrices too) we can play games just like path integration. In space, at least, the transition matrices needed for different actions all have exactly the same eigenvectors, but different (complex) eigenvalues. Hence, path integration can be reduced to successively adding the eigenvalues associated with each action \cite{Yu2020}. This way of thinking unifies path integration with SR-like planning. Interestingly, it also brings different models of path integration into a common framework since, in this case, the eigenvectors are plane waves (not grids as the transitions are unidirectional!) just like those required for VCOs \cite{OKeefe1993, Burgess2009}, and the transition matrix is just like the weight matrices required for CANNs \cite{Burak2006}. 

\end{boxy}

\subsection*{Credit assignment through generalisation and the interplay with striatal RL}

Credit assignment is the attribution of value to state. RL typically assumes that the underlying state-space is fixed, and values are \textit{slowly} assigned to these states. There is no requirement for state representations to be fixed, however; they can change to better represent value. For example, after encountering a goal, GVCs (goal-vector cells) form \cite{Sarel2017} - cells that are active at certain distances and directions from goals (Figure \ref{fig:integrations}d). This can be interpreted as a state representation augmentation (with new cells). Importantly, since GVCs path integrate, once a single GVC forms at a goal, all others can be built for free as the animal navigates the map.

We propose that such compositional representations come `pre-credit assigned', such that they can \textit{immediately} provide a state-space that accurately predicts value (or policy). This can easily be understood in the context of tasks with changing goal locations; pre-learned goal-vector representations can be immediately composed with spatial representations to generate an accurate and flexible representation of any goal state (Figure \ref{fig:integrations}d), and since these GVCs come with value (or policy) already attached, optimal actions can be taken much more rapidly than in the case of traditional RL-learning \footnotemark. The only \textit{online} role of the cognitive map is inferring which pre-learned and pre-credit assigned representations to compose. In the case of goals this is easy - only GVCs are required, though it can be other \textit{local bases} in other circumstances. This is \textbf{credit assignment through generalisation}, and is akin to meta-RL \cite{Wang2018a, Duan2016}, since prior statistical knowledge (e.g. GVCs) can be integrated on-the-fly to solve novel tasks.

\footnotetext{This relates to DR, but here representations are truly compositional - the same GVCs can be used anywhere in space! Additionally, values can be optimal and not just optimal under linear RL assumptions.}

Where do these representations come from in the first place? The cognitive map models suggest that such representations can be learned from statistics of behaviour. Just as OVCs can be learned as reflections of regular movements towards objects \cite{Whittington2020}, GVCs can be learned when behaviour is biased towards goals. In general, to train these `pre-credit assigned' compositional representations, cortex must learn from sequences of behaviour. This suggests an interesting interplay between generalisation and reinforcement learning. Initially, behaviour is generated via classic RL (perhaps in the striatum). Understandably, initial striatal actions will be bad (when encountering entirely novel tasks), but as RL learns good policies, actions will be towards goals. The cortico-hippocampal system can then learn compositional representations of these policies (e.g. GVCs) from the statistics of these sequences. In novel tasks, behaviour can be generated entirely from generalisation, with no need for new striatal RL. This also offers a virtuous cycle, where learned general cortical representations can be provided back to striatum as a state-space for RL, and so on. The notion of offline cortical learning from striatal actions sequences relates to recent machine learning methods in offline RL. Here, sequence models learn the statistics of behavioural sequences from conventional RL algorithms, after which the sequence model can be used for planning \cite{Chen2021a, Janner2021} in a manner analogous to planning by inference. In sum, this proposal offers a novel role of cortical-basal ganglia interaction for constructing RL state-spaces and generalising policies.

\subsubsection*{Replay: offline state-space construction}

If behavioural control in a new world is reduced to a state-space composition problem, it becomes important to construct state-spaces rapidly and accurately, and to store them in memory so they can inform future decisions. To build such memories requires path integration (for example, to ensure the correct goal-vector cell is tied (composed) to the correct location), but to build them quickly this composition should be done as much as possible offline - it should not be the animal's actual location that is integrated.

An appealing substrate for this composition is replay \cite{Foster2006}. For example, when an animal receives reward, it is important that all other states in the environment are aware of their relative location to the reward. Replay can path integrate away from the reward, successively tying (composing) each new goal-vector cell to its respective hippocampal/cortical location (perhaps building landmark cells in hippocampus \cite{Deshmukh2013}; this is a similar mechanism to the simultaneous grid and place cell replay from Evans and Burgess \cite{Evans2019}, but now used to instantiate rewarding policies, instead of ensuring consistency between place and grid representations). Now, should the animal return to a state, that state representation already `knows' about its relation to the reward (Figure \ref{fig:integrations}e). It is no longer necessary to hold all goal locations in mind, as the state-space composition is stored in memory. All heavy computations of building state-spaces take place off-line, and thus the computational burden is reduced for online behaviour. This idea relates to previous ideas from RL that cast replay as a mechanism for optimal credit assignment to existing states \cite{Mattar2018}, or a mechanism for building state-spaces from scratch \cite{Momennejad2017, Sutton2017}. However, in a generalisation framework (outlined in the section above), these two computational processes are subsumed by the single process of composing state-spaces from pre-learnt bases. To test this framework against data, it will be interesting to build a formal understanding of optimal replay patterns under these assumptions. Notably, it will make predictions not only about patterns of hippocampal replay, but also if and when these patterns will align with replay of more abstract representations in entorhinal and frontal cortices \cite{Olafsdottir2016, Kaefer2020}.

\subsection*{When neural representations factorise}

Grid cells were once thought of as representations of space and space alone. By apparently ignoring sensory (or other non-spatial) details of an environment, grid cells were considered a \textit{factorised} representation of space. Similarly, other spatial representations found in entorhinal cortex, such as OVCs and BVCs, are seemingly factorised, since they compositionally augment the entorhinal grid representation to represent different environment configurations. Recent evidence, however, has shown that grid cells warp towards consistently rewarded locations \cite{Boccara2019, Butler2019}. Factorised representations do not warp, since warping is an environment-specific phenomenon; warping around rewards does not transfer to different spatial configurations of rewards.

Why should grid cells warp and sometimes not? There is a computational trade-off between using factorised compositional bases and using bespoke warped representations - specifically, a pressure to generalise versus precisely representing a single task (Figure \ref{fig:integrations}f). With infrequent task switches (that is, repetitively solving a single task), it is more efficient to learn and store a bespoke warped representation (warped since the animal's notion of space becomes warped as it just does a stereotyped looping behaviour in space), as generalisation is not necessary and storing one representation is more efficient than combining many. With regular task switches (such as solving different goal configurations of the same task), the pressure for generalisation is high, and so \textit{compositional} bases are favourable. This idea can be stated succinctly: when the set of tasks that an animal faces is itself factorised, then cellular representations for that task will also be factorised (made up of compositional bases) (Figure \ref{fig:integrations}f). This hypothesis can thus make simple and falsifiable predictions in spatial tasks with environmental rewards. When rewards and space regularly occur in any combination (factorised), both representations of space (grid cells) and reward (reward-vector cells) will exist. By contrast, when rewards and space always occur in the same combination, a bespoke, warped representation will suffice \cite{Boccara2019, Butler2019}.

\section*{Open questions}

\subsection*{The role of time in memory and cognitive maps}

\begin{figure}
\centering
\includegraphics[width=0.95\linewidth]{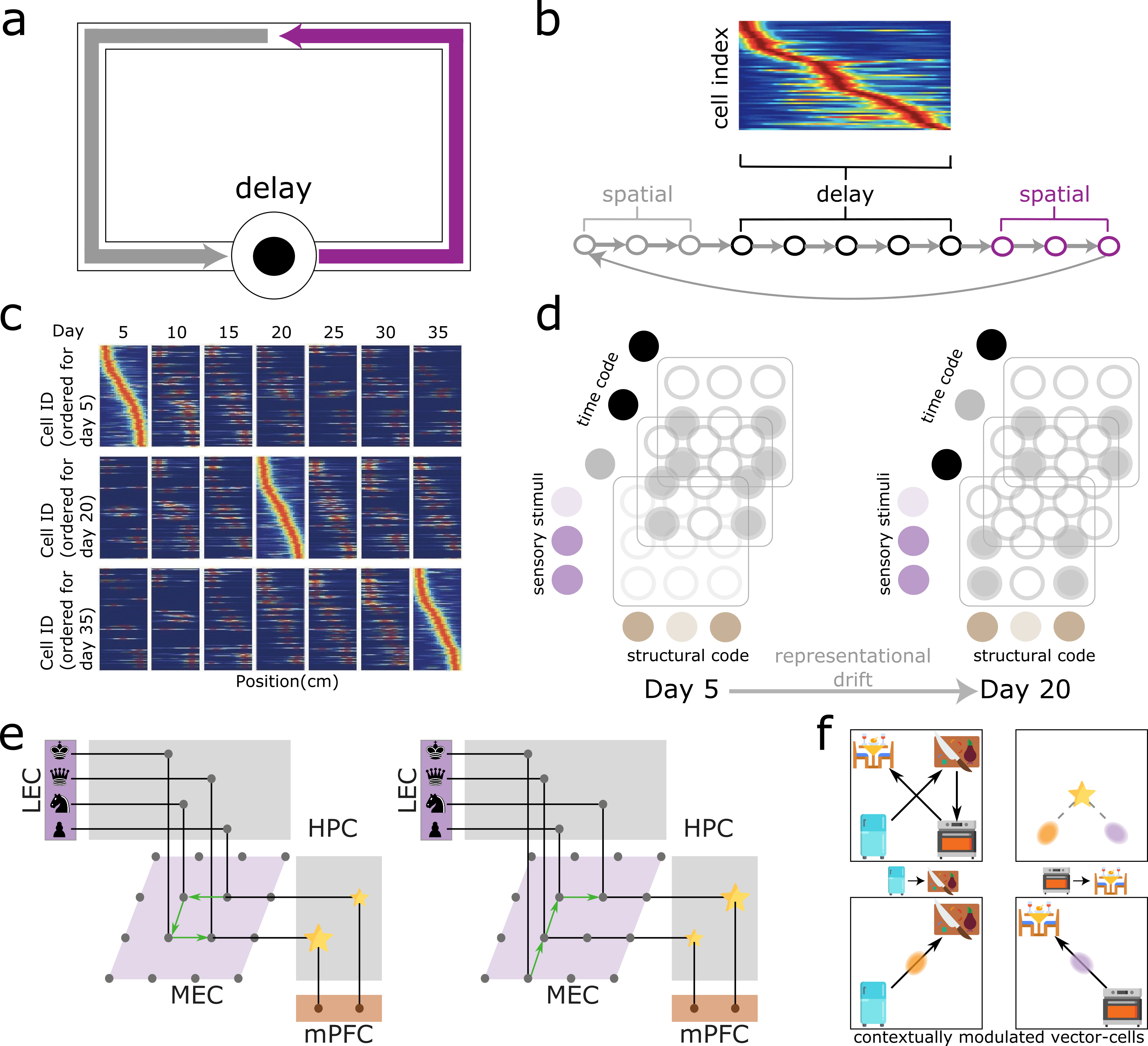}
\caption{\textbf{Representing time and hierarchies of abstraction in cognitive maps.} \textbf{(a-b)} Neuronal representations of time might structure tasks according to `progress' through that task. \textbf{(a)} In a task with a delay period, the full \textbf{(b)} latent state-space of the task includes the delay period, since `progress' through the delay must be represented to predict when the delay period ends: in essence, latent states for predicting the future. Indeed sequences of hippocampal cells fire during the delay period as if they were coding time \cite{Pastalkova2008, MacDonald2011a} (adapted from Salz \textit{et al.} \cite{Salz2016}). \textbf{(c)} Time additionally impacts representations via drift \cite{Ziv2013}. Here, hippocampal (and other) representations slowly change over many days, such that an entirely different representation encodes the same location (adapted from Ziv \textit{et al.} \cite{Ziv2013}). \textbf{(d)} While the mechanism, or function, of drift is unknown, a tantalising possibility, inspired by the reviewed models, is that representational drift is remapping in disguise. In particular, as in TEM, if hippocampal representations reflect a triple conjunction of space, sensory stimuli, and time, then drifting hippocampal representations can parsimoniously be due to changing time representation while space and sensory representations remain the same. Rather than spatial remapping (Figure \ref{fig:tem_smp}b), this is \textit{temporal} remapping. \textbf{(e-f)} Representing hierarchical tasks. \textbf{(e)} Schematic of a \textit{hierarchical} version of TEM, where an additional prefrontal module is included. Should this module represent location in task at an abstracted level (e.g. `just before the over' in a recipe), this abstract and \textit{non-spatial} representation can contextualise the hippocampal-entorhinal system - set goals \textit{in space}. We note that this schematic is a simplification of true neuroanatomy (e.g. mPFC-HC connections may go via reuniens). \textbf{(f)} This predicts novel cell types, such as route-dependent GVCs: representations that point towards goal locations but only at specific points in the task (e.g. only before chopping the vegetables). This is analogous to splitter cells, though these representations can occur anywhere in space, not just at specific points on a T-maze. Icons from \url{https://www.flaticon.com}.}
\label{fig:open_questions}
\end{figure}

The discussion of cognitive map models so far assumes that learned representations remain stable over time. This clearly cannot be the case, since we can remember events at the same place and same conditions but on different days - hippocampal memory representations (e.g. place cells) must be different for different moments in time. An accumulating body of evidence indicates exactly this, with neural representations \textit{drifting} over time and experience (Figure \ref{fig:open_questions}c), challenging traditional notions of engrams and receptive fields \cite{Ziv2013, Driscoll2017, Rule2019, Rubin2015}. 

But how can hippocampus maintain a stable representation of space, if the cellular basis of this representation is drifting over time? Generalisation models offer a natural solution as, here, hippocampal cells bind multiple factors of the input. Only one factor needs to change for the entire hippocampal representation to change (Figure \ref{fig:open_questions}d). If entorhinal cortex could learn abstracted representations of time as well as space, then as time passes the temporal code will progress and the hippocampal code will drift to new cells, but these new cells will only differ in their connections to the entorhinal cells that represent time, not those that represent space (Figure \ref{fig:open_questions}d). Representational drift, in this view, is just hippocampal remapping, but now it is not sensory observations or space that has changed (as in Figure \ref{fig:tem_smp}b), but time instead. A prediction that follows is that the order of drifting cells is not random. 

Hippocampal represents time thought more than just drift. Pure `time cells', for example emerge when rodents are required to stay still, or run on a wheel, for a particular duration of time in a task (Figure \ref{fig:open_questions}a) \cite{Pastalkova2008, MacDonald2011a}. These cells can be easily understood as enabling prediction of when the delay period finishes (Figure \ref{fig:open_questions}b). Crucially, this temporal representation is just one part of a overall map relating experiences to one another. More precisely, during the delay period, while space is not changing, \textit{position in task} is changing. It is this overall task position that cognitive map models suggest is being represented in `time cells'. Indeed, as we have demonstrated in Figure \ref{fig:latent}, neurons representing latent states often appear to be tracking \textit{progress through task}, and in this view, time cells can be interpreted as another instance of this.

Lastly, viewing representational drift through the lens of temporal abstraction might help cognitive maps build abstractions from limited data. In particular, since learning by extracting the statistics of \textit{multiple} different situations is critical for abstraction, drifting representations may allow a \textit{single} environment to instantiate \textit{multiple} representations, appearing (to cortex) as if they were from multiple different environments (Figure \ref{fig:open_questions}d). This process relates to data-augmentation from machine learning \cite{Chen2020}, in which additionally presenting various transformations (such as rotation, cropping, colour distortions, noise addition) of the \textit{same} data dramatically improves representation learning and generalisation, since the abstract quantity of interest is invariant to the various transformations (a rotated dog is still a dog, for example). Similarly, should the structures of interest (within the sequential input data - e.g. loops, space, task) be invariant to drifting hippocampal representations, then abstraction and generalisation could be enhanced. 
In sum, casting representational drift as a form of \textit{temporal abstraction} for generalisation might prove fruitful, and integrate the previously confusing findings of representational drift with existing models of cognitive maps.

\subsection*{Interacting levels of abstraction}

We have shown how models can build abstract representations that generalise over different \textit{sensory} realisations, but the real power of abstractions comes when this process can happen repeatedly, so that abstractions can themselves lead to further abstractions. When we are learning to cook a new recipe, we don't need to relearn the rules of space to find the oven, and when the recipe is learnt it can easily be transferred to kitchens with new spatial layouts.

In the latent state tasks from earlier (e.g. spatial alternation; Figure \ref{fig:latent}), while there was both a task (`go left then right') and space at play, they came in a fixed configuration; the latent space would not have generalised if the T-maze became a W-maze. Cooking recipes in different kitchens is equivalent to a T- to W-maze switch, thus we need something fundamentally new in the models to account for this. One attractive options is for \textit{both} spatial representation and task (\( left \to right \to left \to right \cdots \)) representations need to be separately represented (factorised) so they can be arbitrarily combined (ovens being in different locations in different kitchens). Given enough experiences of different kitchens, this factorisation can emerge from training, and does in many deep learning networks \cite{Higgins2017}. However, using the same tricks as before - i.e. using the hippocampus as a mediator of factorised representations (e.g. space and task) - the required number of recipes and kitchens for training can be dramatically reduced. One intriguing possibility is that the different representations observed in fronto-temporal cortices \cite{Zhou2020a, Zhou2019, Miller2001a, Bernardi2020, Morton2020, Samborska2021} might reflect such a factorisation, with entorhinal representations grounded in interactions with the physical environment, while neurons in PFC representing abstract, task-related invariances, such as `location in task' \cite{Wilson2014a, Schuck2016a, Yu2018, Zhou2020a, Kaefer2020, Samborska2021}.

While a factorisation allows representation of any space-task combination, to actually understand any given space-task combination, these representations must interact. The go-to-oven mPFC representations needs to be linked to the spatial location of the oven, or vector cells pointing towards the oven, in order to navigate to the oven. This linking can occur in exactly the same way as the earlier models suggest - through hippocampal memories (Figure \ref{fig:open_questions}e; we note the mPFC-HPC connection is likely mediated, for example via nucleus reuniens \cite{Ito2015}). Interestingly, though, the very same vector cells can be reused whether it be the oven or the chopping board. This make a prediction - vector cells that are contextually modulated \cite{Miller2001a} depending on `location in task' (Figure \ref{fig:open_questions}f).

Building models of interacting task and spatial representations, with principles of abstraction, generalisation, and path integration, allows neural representations from RL tasks to be understood in the \textit{same} language as space. With emergent task-level (mPFC) representations potentially revealing insights into how cells might represent task structure itself, they will therefore be of interest whenever animals are shaped to perform tasks. 

\subsection*{From sequences to other domains of cognition}

The models we have described translate the problem of building maps into problems of understanding the structure of possible sequences. This raises two interesting points. Firstly, there are many other sequence problems we face that are not traditionally thought of as related to cognitive maps - perhaps these can be understood similarly to space and tasks \cite{Hawkins2019, Lewis2021}. Secondly, sequence problems are not the only cognitive problems we face - how can the understanding we have got form sequences extend to these domains? 

How far can we get with sequence problems? Machine learning has taught us that sequence learners (recurrent neural networks, long short-term memory units \cite{Hochreiter1997}, Transformers \cite{Vaswani2017}) can perform well on a wide variety of tasks including language processing, mathematical understanding, and logic problems \cite{Brown2020,Dosovitskiy2020}. This makes sense since language, mathematics, and formal logic are \textit{sequence} problems where generalisation is key \cite{Dehaene2015, Christiansen1999}. Each comprises of content (words/numbers) combined within different structures (grammatical rules/ mathematical operators), and vice versa. Whilst mathematics and language engage large (and different) cortical territories \cite{Amalric2016}, it is interesting to consider whether the neuronal representations that support these functions might be understood with principles similar to state representation, factorisation, and path integration described in the sections above. For example, mathematical operators, like addition and subtraction, bear similarity to forwards and backwards actions on a line (similar analogies can be made for integration and differentiation). One intriguing finding is that hippocampal and entorhinal cells have fields that respond to unique numbers \cite{Nieder2012}. 

What about non-sequence problems? Much of the neural processing underlying cognitive problems does not seemingly require sequence transitions. For example, understanding a football and the Earth are both spheres does not require learning from sequences; thus, it is not clear if organising principles similar to space play a role in learning these abstractions. Analogies between path integration and understanding spheres, however, can be made. The data-generative factors of a ball - \{sizes, shapes, colours\} - are all examples of variables which can be projected onto a manifold where `actions' such as \texttt{add-red}, \texttt{bigger}, \texttt{remove-red}, then \texttt{smaller} have a meaning (and would take you back to the same sphere!). Indeed modern machine learning methods learn such manifolds from images inputted in no particular sequence \cite{Higgins2017, Higgins2018a}. Some non-sequential problems can also be reformulated sequentially. While an image is not sequential itself, it can be viewed sequentially. In this vein, it is notable that grid-like cells have been observed in monkey \cite{Killian2018} and human \cite{Nau2018, Julian2018} entorhinal cortex during saccades on images. Similarly, when humans view silhouettes of stacked objects, component objects are replayed sequentially \cite{Schwartenbeck2021}. 

\section*{Conclusion}

The hippocampal formation is a poster child for cognitive neuroscience because of its beautifully organised neuronal responses and the profound effects of its damage. However, whilst these experimental findings seem self-explanatory when examined in simple situations like open-field foraging, they have been hard to relate to complex real-world behaviours. Moreover, it has not been clear whether the amazing discoveries gleaned from examining rodents navigating in space might have broader implications for understanding more general cognitive processes. The ideas reviewed in this paper offer concrete and formal methods for addressing these long-standing questions. Excitingly they do this by re-imagining the problem. By asking questions such as`what really is space to the brain' they have been able to make connections between how neurons behave in space, and in many non-spatial tasks. They have provided new computational explanations for how these processes might support behaviour, and for the link between space and memory. Going forward, it is an exciting time to be in the field. These contributions have been made by many researchers across the globe, and have relied on a genuine link between theory and experiment. We believe that this cross-disciplinary collaboration is poised to make big strides in understanding how our brains make sense of the structure of our experience, and use it to construct new flexible behaviours. 

\medskip

\bibliography{references}

\section*{Acknowledgements}

We thank the following funding sources: Sir Henry Wellcome Post-doctoral Fellowship (222817/Z/21/Z) to J.C.R.W.; Wellcome Trust DPhil Scholarship to D.M.; Wellcome Principal Research Fellowship (219525/Z/19/Z), Wellcome Collaborator award (214314/Z/18/Z), and JS McDonnell Foundation award (JSMF220020372) to T.E.J.B.; the Wellcome Centre for Integrative Neuroimaging and Wellcome Centre for Human Neuroimaging are each supported by core funding from the Wellcome Trust (203139/Z/16/Z, 203147/Z/16/Z).

\section*{Author Contributions}

JCRW and TEJB conceptualised manuscript. JCRW and DM performed simulations. JCRW and DM drafted manuscript. JCRW and TEJB edited manuscript with input from all other authors. 

\section*{Data availability}

No data was generated in this perspective.

\section*{Code availability}

Python and Tensorflow code will made available on publication.

\end{document}